\documentclass[12pt,letterpaper]{article}

\usepackage{alltt}
\usepackage{fullpage}
\usepackage{multirow}
\usepackage{rotating}
\usepackage[english]{babel}
\usepackage{color}
\usepackage{colortbl}
\usepackage{natbib}
\usepackage{float}
\usepackage{url}
\usepackage{graphics}
\usepackage{amsmath}
\usepackage{amsthm}
\usepackage{amssymb}
\usepackage[all]{xy}
\usepackage{booktabs}
\usepackage{framed}
\usepackage{dcolumn}
\usepackage[T1]{fontenc}
\usepackage{rotating}
\usepackage{caption}
\usepackage{epstopdf}

\newcolumntype{.}{D{.}{.}{-1}}

\floatstyle{plaintop}
\restylefloat{table}

\makeatletter
\def\maxwidth{ %
  \ifdim\Gin@nat@width>\linewidth
    \linewidth
  \else
    \Gin@nat@width
  \fi
}
\makeatother

\IfFileExists{upquote.sty}{\usepackage{upquote}}{}
\definecolor{fgcolor}{rgb}{0.2, 0.2, 0.2}

\makeatletter
 {\par\unskip\endMakeFramed%
 \at@end@of@kframe}
\makeatother

\definecolor{shadecolor}{rgb}{.97, .97, .97}
\definecolor{messagecolor}{rgb}{0, 0, 0}
\definecolor{warningcolor}{rgb}{1, 0, 1}
\definecolor{errorcolor}{rgb}{1, 0, 0}

\renewcommand{\section}[1]{%
\bigskip
\begin{center}
\begin{Large}
\normalfont\scshape #1
\medskip
\end{Large}
\end{center}}

\renewcommand{\subsection}[1]{%
\bigskip
\begin{center}
\begin{large}
\normalfont\itshape #1
\end{large}
\end{center}}

\renewcommand{\subsubsection}[1]{%
\vspace{2ex}
\noindent
\textit{#1.}---}

\renewcommand{\tableofcontents}{}

\DeclareSymbolFont{letters}{OML}{ztmcm}{m}{it}
\DeclareSymbolFontAlphabet{\mathnormal}{letters}

\def\title{Inferring Heterogeneous Evolutionary Processes Through Time: from sequence substitution to phylogeography}

\definecolor{snow3}{rgb}{0.8,0.78,0.78}
\definecolor{Gray}{gray}{0.9}

\theoremstyle{plain}

  \theoremstyle{definition}
  
  \theoremstyle{remark}

\providecommand{\definitionname}{Definition}
\providecommand{\remarkname}{Remark}
\providecommand{\theoremname}{Theorem}

\providecommand{\tabularnewline}{\\}
\floatstyle{ruled}
\newfloat{algorithm}{tbp}{loa}
\providecommand{\algorithmname}{Algorithm}
\floatname{algorithm}{\protect\algorithmname}

\raggedright
\begin{document}

\begin{flushright}
Version dated: \today
\end{flushright}
\bigskip

% running head
\noindent Epoch substitution models

\bigskip
\medskip
\begin{center}

\noindent{\Large \bf 
\title
}
\bigskip

\noindent {\normalsize \sc Filip Bielejec$^1$, Philippe Lemey$^1$, Guy Baele$^1$, Andrew Rambaut$^{2,3}$ and Marc A.~Suchard$^{4,5}$ }\\
\noindent {\small \it 
$^1$Department of Microbiology and Immunology, Rega Institute, KU Leuven, Leuven, Belgium;\\
$^2$Institute of Evolutionary Biology, University of Edinburgh, Edinburgh, United Kingdom \\
$^3$Fogarty International Center, National Institutes of Health, Bethesda, MD, USA \\
$^4$Departments of Biomathematics and Human Genetics, David Geffen School of Medicine at UCLA, University of California, Los Angeles, CA, 90095, USA.\\
$^5$Department of Biostatistics, UCLA Fielding School of Public Health, University of California, Los Angeles, CA, 90095, USA.}\\
\end{center}
\medskip
\noindent{\bf Corresponding author:} Filip Bielejec, Department of Microbiology and Immunology, Rega Institute, KU Leuven, Leuven, Belgium; E-mail: filip.bielejec(at)rega.kuleuven.be\\

%\vspace{1in}

\newpage

\section{Abstract}

Molecular phylogenetic and phylogeographic reconstructions generally assume time-homogeneous substitution processes.
Motivated by computational convenience, this assumption sacrifices biological realism and offers little opportunity to uncover the temporal dynamics in evolutionary histories.
Here, we extend and generalize an evolutionary approach that relaxes the time-homogeneous process assumption by allowing the specification of different infinitesimal substitution rate matrices across different time intervals, called epochs, along the evolutionary history.
% TODO: add what else is new, ie we don't assume a known phylogeny and are not branch specific
We focus on an epoch model implementation in a Bayesian inference framework that offers great modeling flexibility in drawing inference about any discrete data type characterized as a continuous-time Markov chain, including phylogeographic traits.
To alleviate the computational burden that the additional temporal heterogeneity imposes, we adopt a massively parallel approach that achieves both fine- and coarse-grain parallelization of the computations across branches that accommodate epoch transitions, making extensive use of graphics processing units.
\par
Through synthetic examples, we assess model performance in recovering evolutionary parameters from data generated according to different evolutionary scenarios that comprise different numbers of epochs for both nucleotide and codon substitution processes. 
We illustrate the usefulness of our inference framework in two different applications to empirical data sets: the selection dynamics on within-host HIV populations throughout infection and the seasonality of global influenza circulation.
In both cases, our epoch model captures key features of temporal heterogeneity that remained difficult to test using \textit{ad hoc} procedures.

\noindent (Keywords: Phylogenetics, Phylogeography, Bayesian inference, Epoch Model, BEAST, BEAGLE
%, Unwieldy Titles
)\\

%\vspace{1.5in}
\newpage

\section{Introduction}

Molecular phylogenetic models typically consider sequence evolution as a continuous-time Markov chain (CTMC) that operates along the branches of a bifurcating tree.
As a description of the character substitution process, CTMCs take their values from a finite set of discrete states called the state space and satisfy the Markov property.
Current models are not limited to nucleotide or amino acid data, and frequently accommodate large state spaces, such as codon substitution models \citep{Goldman1994, Muse1994}, and generalize to many discrete data type including phylogeographic traits \citep{Lemey2009}, where not only the state space can be large but also the underlying substitution rate matrix may be asymmetric.
The Markov property ensures that the process is memoryless, implying that the conditional probability distribution of future states only depends upon the present state, and not on the preceding sequence of events.
CTMCs are characterized by matrices of infinitesimal rates that quantify the probabilities of exchanging discrete characters in an infinitely small time interval. 

Phylogenetic inference often resorts to substitution processes that are stationary, homogeneous and reversible.
Stationarity dictates that the process is at equilibrium, such that the frequency distribution of realized states remains constant over the course of evolution. 
Homogeneity ensures that the process is constant in pattern throughout evolutionary history, thereby treating evolution as a lineage- or time-independent process.
This implies that non-stationarity induces inhomogeneity, as the process of evolution depends upon the equilibrium frequencies.
However, a process can be stationary but not homogeneous, e.g. through the specification of different instantaneous rate parameters for different parts in the tree.
Finally, reversibility is a frequently applied restriction to the rate matrix describing molecular evolutionary processes that leads to a reduced number of free parameters.
Collectively, these restrictive assumptions make strong abstraction of the underlying substitution process to ease mathematical and computational tractability.

Recently, substantial work has strived to relax the standard assumptions in CTMC processes, in order to uncover more complex evolutionary processes and assess their impact on phylogenetic reconstruction. 
To accommodate non-stationarity, \citet{Yang1995}, \citet{Galtier1998} and \citet{Galtier1999}, for example, have proposed models that allow the nucleotide composition to vary over the tree.
Although this includes general treatments involving separate composition parameters for each tree branch, large trees will inevitably lead to over-parameterized models.
To address this problem, \citet{Foster2004} has developed an approach that maps a restricted, but fixed number of nucleotide composition vectors with estimable frequencies to the tree. 
Set in a Bayesian framework, this approach also integrates over the tree topology and finds improved posterior support estimates for topologies in examples with compositional heterogeneity, as opposed to inference under a stationary model that would suffer from attraction artifacts due to similar compositional biases. %PL: try to work in \cite{Gowri-Shankar2007} here?
Further developments have uncoupled compositional shifts from particular nodes in the tree while estimating the total number of events of compositional drift distributed across the tree using a compound Poisson process \citep{Blanquart2006}. 
\cite{Blanquart2008} further such approaches for amino acid evolution in conjunction with models that take into account site-specific  substitution patterns induced by protein structure and function.
Models also exist to tackle inhomogeneity in the instantaneous rates of change rather than perturbing stationarity, such as a codon substitution model that allows for the non-synonymous to synonymous substitution rate ratio ($d_N/d_S = \omega$) to vary among branches while keeping the codon equilibrium frequencies constant \citep{Yang1998}.
In general, tree-based modeling of heterogeneity in the pattern of evolution typically finds its use when applied to widely divergent taxa representing relatively rich speciation histories and possibly involving lineage-specific adaptation.

\par  % MAS: new paragraph to highlight change

However, a change in the evolutionary process may also apply to an entire population at a particular point in time, in which case the evolutionary shift simultaneously cuts across all lineages at that time point in the underlying genealogy.
\citet{Goode2008} first consider such a scenario; they develop
an extension of a codon substitution model with discrete site classes that allows for a time-specific change in $\omega$ and in the transition/transversion rate ratio ($\kappa$), and assign prior probabilities that sites belong to a particular class.
Specifying such change-points requires trees measured in time and so \citet{Goode2008} adopt a strict molecular clock model on a fixed tree topology in order to apply the model % AR: changed this because 2 consecutive sentences starting with 'because'  
to HIV envelope sequences sampled from a single patient over a period of three years.  
Because the rapidly evolving virus population accumulates significant substitutions over such a short time-scale, one can estimate the rate of evolution by incorporating the sampling dates of the sequences (the `dated tip' model, \citet{Rambaut2000}).
The authors demonstrate that many sites classified as neutral or under positive selection before therapy appear to be under strong negative selection upon treatment initiation.

We build upon the approach of \citet{Goode2008} in several important ways.
By implementing a similar model of time-specific evolutionary changes in the Bayesian Evolutionary Analysis by Sampling Trees (BEAST) software package \citep{Drummond2012}, 
we connect the epoch models to different relaxed clock models that often provide a more realistic description of the tempo of evolution \citep{Drummond2006, Drummond2010}.
More importantly, we generalize the epoch model to any finite discrete data type and any number of transition times.
The former is critical to accommodate discrete phylogeographic inference \citep{Lemey2009}, for which \citet{Bahl14112011} recently demonstrate the need to incorporate time-specific migration rates.
Our Bayesian approach also does not condition on a fixed tree topology but averages over all plausible evolutionary histories.  This integration naturally accounts for uncertainty in the tree and in how the epoch transition times translate to varying branch-specific change points.
Jointly estimating the epoch-associated rate matrices and the unknown evolutionary history also ensures that we can fully exploit our Bayesian phylogeographic (or discrete trait evolutionary) inference, which explicitly connects sequence evolution to the trait diffusion process \citep{Lemey2009}.
Finally, our implementation also allows us to make use of recent marginal likelihood estimators to assess model fit for different epoch parametrizations \citep{Baele07032012}.

Recent advances in sequencing technology confront statistical phylogenetics with escalating data demands. 
Because the number of possible tree topologies grows explosively with the number of sequences, integrating over the evolutionary history becomes a computationally daunting task.
To make matters worse, considerable computational effort is already required to evaluate each history, in particular for complex evolutionary models with large state spaces.
For this reason, codon substitution models, for example, have largely been neglected in Bayesian phylogenetic inference. 
Recently, however, new algorithms have been developed to exploit massive parallelization on graphics processing units (GPUs), offering dramatic speed increases for statistical inference under complex evolutionary models \citep{Suchard2009,Baele12062013}.
By partitioning the time component into discrete intervals, the epoch model further adds to the computational burden, but it also represents an opportunity to exploit massive parallel computation \citep{Suchard2009,suchard2010some}. 
To apply the epoch substitution heterogeneity in conjunction with large-state space models to large data sets, we implement our model as part of the Broad-platform Evolutionary Analysis General Likelihood Evaluator (BEAGLE) library for evaluating the likelihood of sequence evolution on trees \citep{ayres2012beagle}, taking the effort to accommodate multiple scales of parallelization to keep computation time manageable. 

Following \citet{Goode2008}, we mainly focus on rapidly evolving populations for which significant divergence accumulates between sequences sampled at different time points, both from the simulation perspective as for the real data sets.
Using a simulation study we demonstrate that our model can be fit to complex data sets and consistently captures time-specific evolutionary parameters, but is not restricted to time-stamped data.
We further demonstrate the use of our model by examining two real-life examples. 
The first application tests and quantifies changes in $d_N/d_S$ associated with HIV disease progression in several different patients. 
This analysis aims at testing different hypotheses explaining why viral divergence stabilizes close to disease onset in HIV infection \citep{Shankarappa99}. 
The second application employs epoch modeling to accommodate seasonality in the inference of global influenza dispersal dynamics. 
To demonstrate the scalability of the epoch model implementation in BEAST/BEAGLE, we highlight the important, but problem-specific, speed increases a parallel implementation has to offer. 

\newpage

%%%%%%%%%%%%%%%
%---METHODS---%
%%%%%%%%%%%%%%%

\section{Methods}

Continuous-time Markov chain substitution models provide the cornerstone of computational phylogenetics.  Given a discrete trait obtaining $K$ distinct states, a $K \times K$ infinitesimal rate matrix $\mathbf{Q}$ characterizes its CTMC. Matrix $\mathbf{Q}$ contains instantaneous transition rates $q_{ij} \ge 0$ for $i \neq j$ and
satisfies $\mathbf{Q} \mathbf{1} = \mathbf{0}$, where $\mathbf{1}$ and $\mathbf{0}$ are $K \times 1$ column vectors.

From the rate matrix $\mathbf{Q}$, a stochastic matrix $\mathbf{P}$ is computed over time $t \geq 0$ via matrix exponentiation $\mathbf{P}(t)=
\exp(t\mathbf{Q})
=\underset{j=0}{\overset{\infty}{\sum}}
(t\mathbf{Q})^{j} / j!$. For an overview of methods to numerically approximate a matrix exponential, we refer to \citet{Moler78nineteendubious}.
Drawing realizations with probabilities defined by $\mathbf{P}$ gives rise to 
a stochastic process $\left\{ X(t):\; t\geq0\right\}$ satisfying the Markov property, such that for every $n\geq 0$, given the time points $0\leq t_{0}\leq t_{1}<\ldots<t_{n}\leq t_{n+1}$ and discrete states $i_{0},i_{1}, \ldots, i_{n},i_{n+1}$ it holds that $P\left\{ X(t_{n+1})=i_{n+1}\mid X(t_{n})=i_{n},\ldots, X(t_{0})=i_{0}\right\} =P\left\{ X(t_{n+1})=i_{n+1}\mid X(t_{n})=i_{n}\right\} $. 
In general, one refers to the elements of $\mathbf{P}$ as finite-time transition probabilities between the $K$ discrete state-space elements. Let us denote a transition probability between two states $i$ and $j$ over time $u$ to $t+u$ by

\begin{equation}
p_{ij}\left(u,t+u\right)=P\left\{ X(t+u)=j\mid X(u)=i\right\} .
\end{equation}

In the phylogenetic setting, researchers often further constrain these processes 
to be time-homogeneous and time-reversible. Time-homogeneity mandates that  transition probabilities depend only on the difference $t$ between times $u$ and $t + u$,
\begin{equation}
p_{ij}\left(u,t+u\right)= p_{ij}\left(0,t\right) \equiv p_{ij}\left(t\right) .
\end{equation}
Time-reversible CTMCs satisfy detailed balance, such that
$
\pi_{i}p_{ij}(t)=\pi_{j}p_{ji}(t)
%,
$
for all $i$, $j$ and $t$, where $\pi_j = p_{ij}(\infty)$ for all $j$ return the stationary distribution of the CTMC and are independent of starting state $i$.  Finally, common practice in phylogenetics reparametrizes the elements of $\mathbf{Q}$ into relative rates through the constraint $\sum_i \pi_i q_{ii} = -1$ and then, for studies involving phylogenies set in calendar time, multiples $\mathbf{Q}$ by a rate scalar $r$ to form the argument to $\mathbf{P}(t)$.  In this case, we define
\begin{equation}
p_{ij}(u,t+u,r) 
=
\left\{
\exp(
r  t  \mathbf{Q})
\right\}_{ij}
,
\label{eq:makeprobs}
\end{equation}
where $\{ \cdot \}_{ij}$ extracts the $ij$-th element.

\newcommand{\tmax}{t_{\mbox{\tiny max}}}

\citet{Felsenstein1981} provides an efficient algorithm for computing the likelihood of a phylogenetic tree $\mathbf{F}$ given discrete traits and the finite-time transition probabilities along each branch of $\mathbf{F}$.
Label the nodes $x_{1},\ldots,x_{2N-1}$ in an $N$-tipped $\mathbf{F}$ set in calendar time%
.
Now, consider a trio of nodes $u$, $v$ and $w$ where node $u$ lies at time $t_{u}$ in the past and is parent to both nodes $v$ and $w$, at times $t_v$ and $t_w$, respectively, in $\mathbf{F}$.
Then we imagine that an unobserved discrete trait $i$ evolves independently into $j$ at node $v$ over the time interval $\left[t_{u},t_{v}\right]$ with rate scalar $r_{v}$ and into $k$ at node $w$ over $\left[t_{u}, t_{w}\right]$ with rate scalar $r_{w}$.

Visiting all the nodes in post-order fashion, we can integrate out these unobserved traits, calculating successive contributions to the partial likelihood for each node via
\begin{equation}
L_{x_{u}}(i)=
\left[\underset{j}{\sum}p_{ij}(t_{u},t_{v},r_{v})L_{x_{v}}(j)\right]
\times
\left[\underset{k}{\sum}p_{ik}(t_{u},t_{w},r_{w})L_{x_{w}}(k)\right].
\label{eq:partial_likelihood}
\end{equation}

For tip nodes in $\mathbf{F}$, we assign $L_{x_{u}}(i)$ to either $0$ or $1$ depending on whether trait $i$ is (partially) observed or not.  Finally, the full likelihood of $\mathbf{F}$ becomes $\sum_{i} L_{x_{2n-1}}(i) \pi_{i}$, where $x_{2N-1}$ is the root node.  For multiple traits or sequences of length $L$ and for among-site rate mixtures with $C$ categories, one assumes conditional independence across sites and rate categories and simply aggregates site-category contributions.  The serial computational order of this recursion is ${\cal O}( K^2 \times N \times C \times L)$.  

Central to the recursive tree-pruning in Equation (\ref{eq:partial_likelihood}) is specification of the branch-specific transition probabilities
$\mathbf{P}(t_u, t_v, r_v) = \left\{
p_{ij}(t_u,t_v,r_v)
\right\}$ for all $i,j$.  These are commonly homogeneous and conveniently collapse into
functions of just the branch length $t_u - t_v$ instead of the more elaborate starting and ending time.
A strict molecular clock assumption specifies that all $r_u$ are equal, but this is not a necessary restriction of our model because we can allow for the introduction of lineage-specific rate variation in addition to the time-inhomogeneity in substitution processes that we tackle next. 

\subsection{Relaxing time-homogeneity}

The epoch model finds its use in situations where the usual time-homogeneity assumption is violated in specifying $\mathbf{P}(t_u, t_v, r_v)$. 
To model inhomogeneity in process through time, we assume that there exist $S$ unique substitution processes characterized through rate matrices $\mathbf{Q}_s$ for $s = 1,\ldots,S$, and that, at any given point in time, one of these processes is active across all of the extant lineages in $\mathbf{F}$. 
We then model how the active process changes over time via a change-point process with $M + 1$ ordered boundaries at times 
$-\infty = T_0 < T_1 < \cdots < T_{M-1} < T_{M} = \tmax$, where $\tmax$ is the time of the most recently observed tip in $\mathbf{F}$, 
and $M$ indicator functions $\phi_m\ \in\ \{1,\ldots,S\}$ that identify which $\mathbf{Q}_{\phi_m}$ is active during the time epoch 
% $\left(T_{m-1}, T_{m}\right)$.
$\left[T_{m-1}, T_{m}\right]$.
For the examples in this paper, we assume that $S$, $M$ and $(T_m, \phi_m)$ for all $m$ are fixed through marked biological constraints.  

To compute $\mathbf{P}(t_u, t_v, r_v)$ for each branch in $\mathbf{F}$ under this change-point process, we return to the Markov property of CTMCs that says one only needs to keep track of the immediate past in determining transition probabilities for the future.  
This greatly simplifies and regularizes computation, allowing for its parallelization.  Assume $t_u$ lies in epoch $m^{\prime}$ and $t_v$ lies in epoch $m^{\prime\prime}$.  If $m^{\prime} = m^{\prime\prime}$, then no new work is necessary.  We compute these transition probabilities directly via Equation (\ref{eq:makeprobs}) from an eigen-decomposition of $\mathbf{Q}_{m^{\prime}}$; \citet{Suchard2009} describe parallelization of this work across branches and rate categories.
%PL: would it be appropriate to add to the previous sentence a reference to Appendix A
On the other hand, if $m^{\prime} \neq m^{\prime\prime}$,
the branch traverses $m^{\prime\prime} - m^{\prime}$ epoch boundaries at which times $\mathbf{Q}$ changes. To handle these discontinuities, we imagine a data augmentation procedure to break the inhomogeneous process into a conditionally independent series of homogeneous processes and then integrate out the augmented data.

%---CONVOLUTION ILLUSTRATION---%
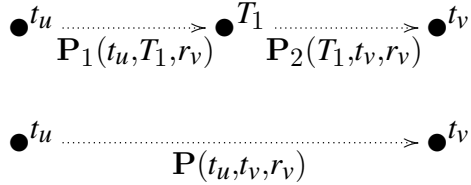
\begin{figure}[H]
\begin{center}
\begingroup
\everymath{\displaystyle}
{\Large
\begin{displaymath} % 
    \xymatrix{ 
    \bullet^{t_u} \ar@{.>}[rr]_{\mathbf{P}_{1}(t_{u},T_{1},r_v)} && \bullet^{T_1} \ar@{.>}[rr]_{\mathbf{P}_{2}(T_{1},t_{v},r_v)} && \bullet^{t_v} \\
    \bullet^{t_u} \ar@{.>}[rrrr]_{\mathbf{P}(t_{u},t_{v},r_v)} & & & & \bullet^{t_v}
    } % 
\end{displaymath}
}
\endgroup
\end{center}
\caption{{ \footnotesize {\bf Collapsing branches.} Transition probability matrix $\mathbf{P}(t_u,t_v,r_v)$ governs the inhomogeneous substitution process along a branch from time $t_u$ to $t_v$ and is the matrix-product of transition matrices $\mathbf{P}_{1}(t_u,T_1,r_v)$ and $\mathbf{P}_{2}(T_1,t_v,r_v)$, where $T_1$ is the epoch change-point time between homogeneous processes 1 and 2.  We assume rate scalar $r_v$ remains constant along the entire branch.
}}
\label{fig:collapsing}
\end{figure}

Figure \ref{fig:collapsing} illustrates this action for a branch that spans a single boundary at $T_1$ with $\mathbf{Q}_1$ governing the process before the boundary and $\mathbf{Q}_2$ after the boundary. Letting $X(T_1) = k$ represent the augmented state of the stochastic process at the boundary, we compute
\begin{align}
p_{ij}(t_u, t_v, r_v)
&= \sum_{k} 
\left\{
\exp[ r_v ( T_1 - t_u ) \mathbf{Q}_1]
\right\}_{ik}
\times
\left\{
\exp[ r_v ( t_v - T_1 ) \mathbf{Q}_2]
\right\}_{kj} , 
\end{align}
for all $i,j$, or equivalently in compact matrix form
\begin{align}
\mathbf{P}(t_u, t_v, r_v)
&= 
\exp[ r_v ( T_1 - t_u ) \mathbf{Q}_1]
\times
\exp[ r_v ( t_v - T_1 ) \mathbf{Q}_2] \nonumber \\
&= \mathbf{P}_{1}(t_u, T_1, r_v) \times \mathbf{P}_{2}(T_1, t_v, r_v) ,
\label{matrixProduct}
\end{align}
where 
$\mathbf{P}_{1}(t_u, T_1, r_v)$ and $\mathbf{P}_{2}(T_1, t_v, r_v)$ are shorthand notation used in the figure and again in the next section where it is clear we are considering substitution models for neighboring epochs.

\par

We colloquially refer to the action of Equation (\ref{matrixProduct}) as a transition probability matrix convolution to remind the reader that we are integrating out an unobserved state in the middle, but in a strict sense, this action is simply matrix multiplication and exemplifies a Chapman-Kolmogorov equation \citep[see, e.g.,][]{Feller1}, %(e.g. \citet{Feller1}), 
% describing short time evolution of a stochastic process that realizes discrete outcomes. 
stating that every stochastic process emitting discrete outcomes as a function of time can be marginalized over one of its variables.

For general $m^{\prime} \neq m^{\prime\prime}$, we arrive at
\begin{align}
\mathbf{P}(t_u, t_v, r_v)
= 
& \exp[ r_v ( T_{m^{\prime}} - t_u ) \mathbf{Q}_{\phi_{m^{\prime}}}] \ \times %\nonumber \\
%& 
\prod_{
\nu = m^{\prime} + 1
}^{
m^{\prime\prime} - 1
}
\exp[ r_v ( T_{\nu} - t_{\nu - 1} ) \mathbf{Q}_{\phi_{\nu}}]
 \nonumber \\
&
\quad \times
\exp[ r_v ( t_v - T_{m^{\prime\prime}} ) \mathbf{Q}_{\phi_{m^{\prime\prime}}}] 
.
\label{eq:convolution}
\end{align}

Each matrix convolution in Equation (\ref{eq:convolution}) is ${\cal O}(K^3)$, potentially commanding a high computational burden compared to the likelihood recursion when $K$ is large and many branches in $\mathbf{F}$ transect multiple boundaries.  
Fortunately, these operations are very regular and both fine- and coarse-grain parallelization offers a solution to the computational burden.

We implement our epoch model in the BEAGLE library \citep{ayres2012beagle} interfaced through the BEAST software package \citep{Drummond2012}.  
Our BEAST/BEAGLE implementation supports extensive parallel computing on state-of-the-art computer hardware, including GPUs through the Compute Unified Device Architecture (CUDA) framework \citep{Nickolls2008}.
Appendix A describes our implementation in BEAGLE to achieve efficient fine-scale parallelization, i.e.\ the type of parallelism where many individual threads are responsible for executing small, in the sense of computational complexity and time required for completion, portions of full task as opposed to coarse-grain parallelism - where a handful of threads is responsible for executing relatively complex and time consuming tasks.

\subsection{Coarse-grain parallel implementation}

We leverage coarse-grain parallelism by collecting sets of independent matrix update and convolution operations across branches through a post-order tree traversal and executing the same operations simultaneously on the GPU.
To this end, our BEAST/BEAGLE implementation includes a front-end routine that keeps track of branches spanning multiple epochs, the time they spend in each epoch and the order of convolutions for asynchronous dispatch.
This routine interfaces with BEAGLE via its application programming interface (API), making calls to the original parallel BEAGLE kernel \emph{UpdateTransitionMatrices} to calculate finite-time transition matrices and the new \emph{ConvolveTransitionMatrices} kernel to multiply two transition matrices into a third product matrix.
Although BEAGLE focuses on evaluating likelihoods of character evolution on phylogenies, BEAGLE has no notion of a tree structure and operates directly on the data, i.e.~indexed, column-major flattened out matrices residing within BEAGLE-managed memory, called buffers.
\par
For each branch, the post-order traversal computes a set of weights that measure the times that each of the substitution processes is active on that branch, summing to the total branch length. 
If a branch's entire weight is within a single interval, no extra work is needed and the finite time transition probabilities are calculated from the eigen decomposition of the single rate matrix active in that interval.
If the branch spans over multiple epochs, partial transition matrices are formed from the weights and eigen decompositions specific to the corresponding intervals.
For those weights that are non-zero in the same time interval, we first push their corresponding buffers to an update queue.  After the update queue grows full, it gets dispatched for parallel execution, and then we begin to place convolve operations in an convolve queue. 

Our implementation keeps track of two work queues, grouping update and convolve operations in the correct order for later execution on the device. 
A single work item in the queue responsible for updating transition matrices will include one buffer that needs to be populated by finite-time transition probabilities, whereas a work item in the queue that convolves transition matrices will have two input buffers and one output buffer for storing the result of the matrix-matrix multiplication. 
Both work queues utilize a first in, first out (FIFO) abstraction 
to guarantee that the input buffers in the convolution are non-empty, having  been populated by the update operation. 
The buffers for which tasks have been completed are then returned to the pool of extra buffers for further use.

\begin{figure}[H]
\begin{center}
\includegraphics[scale=0.5]{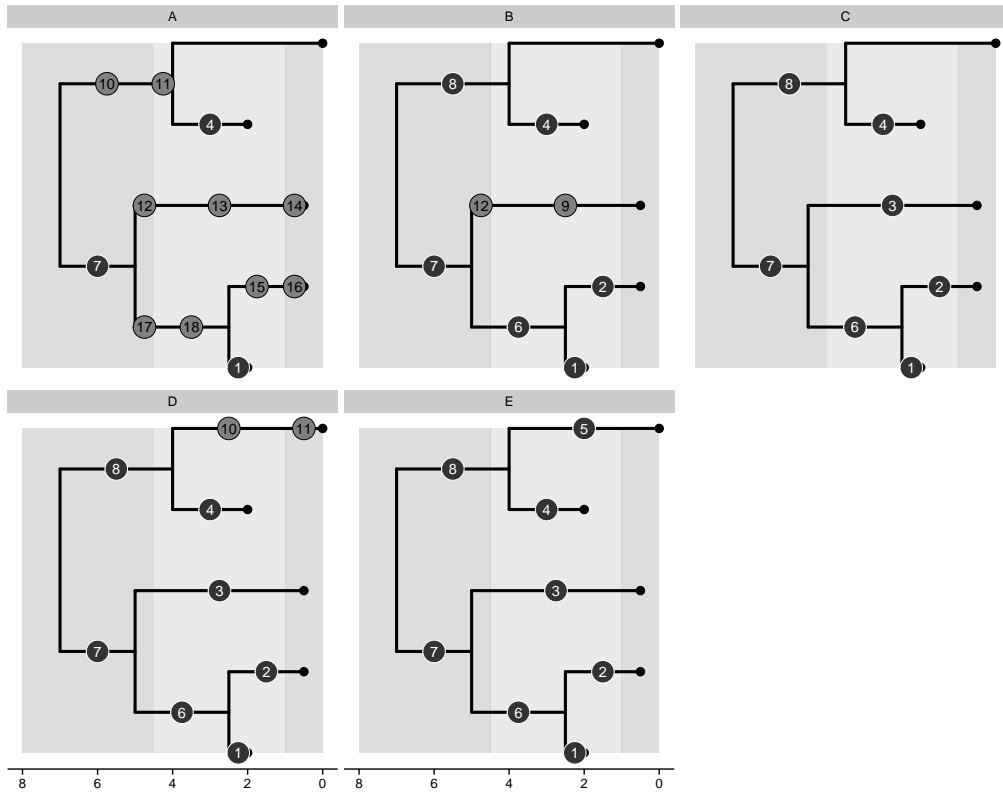} 
\end{center}
\caption{{ \footnotesize {\bf A small-scale example of the update convolve routine for a 5-taxon tree.} 
In this example, transition times are set at $t_{1} = 1.0$ and $t_{2} = 4.5$, creating three respective epochs indicated by alternating background.
Black dots represent transition probability buffers for particular branches, grey dots denote extra buffers allocated for partial transition matrices.
We allocate 10 extra buffers.  
Consecutive panels illustrate the order of update and convolve steps in the routine.
Update step in A adds buffers to the update work queue. 
In B and C, the routine starts adding buffers to the convolve queue in a stepwise fashion, dictated by the tree traversal, freeing extra buffers whenever the queue is dispatched for execution. 
In D, the routine uses freed buffers for a new update step.
In E, the parallel convolve queue is used to compute the final transition matrix stored in buffer 5, concluding the work.
} }
\label{fig:UpdateConvolveExample}
\end{figure}

We coordinate the routine workload via asynchronous work queues because of the limited amount of resources available on the device. A single instance of BEAGLE is invoked with a fixed number of allocated buffers, and at present, it is not dynamically increased.
In the Bayesian inference framework utilized by BEAST, the number of convolution operations becomes random and may well exceed the size set for our buffers.
The amount of required buffers further increases by the need to perform convolution operations on a single branch sequentially  
(see Equation \ref{eq:convolution}) and partial results need to be stored before their dependencies are processed and the work can continue. 
When creating a BEAGLE instance, we allocate a number of extra buffers in addition to the buffers that hold transition matrices for each branch. 
Those buffers are then used to store partial transition matrices and results of matrix-matrix multiplications operations. 
Our routine performs operations in batches of equal or smaller size compared to the number of allocated buffers, by pushing those queues to the GPU for parallel execution once the routine has no more buffers available or all work items are queued. Figure \ref{fig:UpdateConvolveExample} presents an example of update-convolve routine using an asynchronous queue. 

%---UPDATE-CONVOLVE INFOGRAPHIC---%

\clearpage

\section{Results}

%%%%%%%%%%%%%%%%%%%
%---SIMULATIONS---%
%%%%%%%%%%%%%%%%%%%
\subsection{Performance assessment using simulation}

To evaluate the performance of the epoch model, we conduct a simulation study, in which replicate data are generated along an evolutionary history inferred from a real 
data set with samples collected at different points in time.
Specifically, we use a maximum clade credibility (MCC) tree summarizing a Bayesian phylogenetic inference of human influenza A hemagglutinin gene sequences sampled through different epidemic seasons \citep{Drummond2010}. 

%---TOPOLOGY FOR SIMULATIONS---%
\begin{figure}[H]
\begin{center}
% \definecolor{shadecolor}{rgb}{0.969, 0.969, 0.969}
\color{fgcolor}
\includegraphics[scale=0.5]{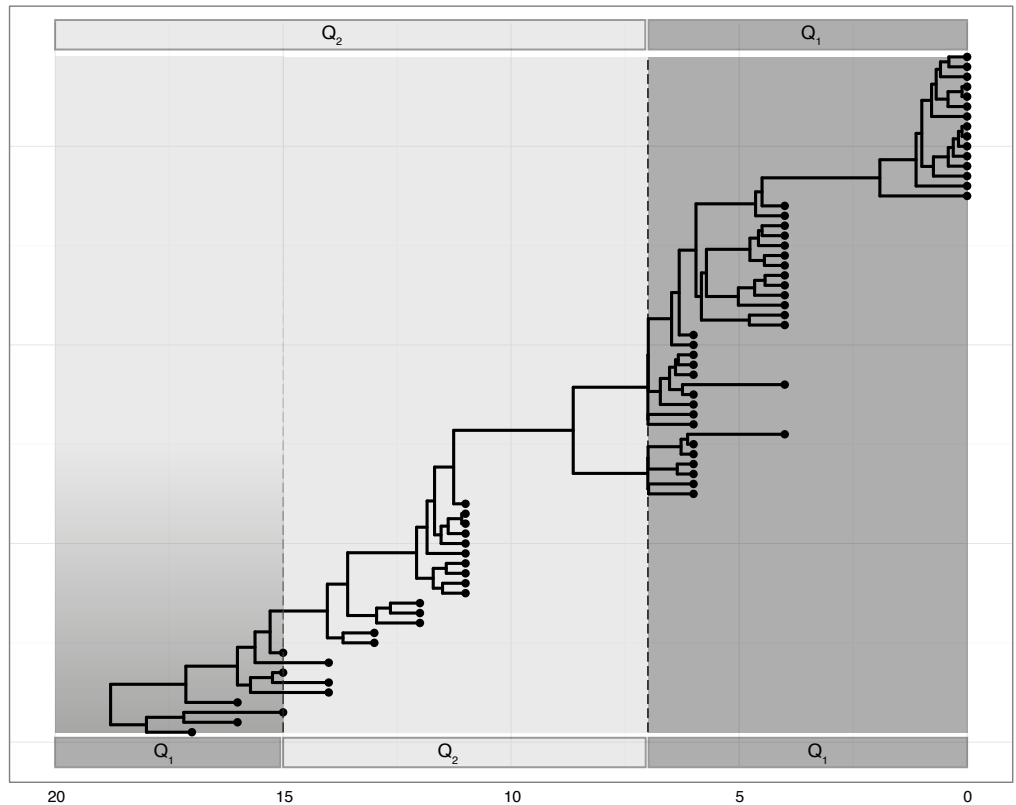} 
\end{center}
\caption{{ \footnotesize {\bf  Epoch simulation scenarios on an influenza A maximum clade credibility tree topology.} 
In the two-epoch example illustrated at the top, the transition time is set at $t_{1}=7$, creating two epochs with substitution processes governed by infinitesimal rate matrices $\mathbf{Q_{1}}$ and $\mathbf{Q_{2}}$ respectively, separated by the light and dark grey areas and the dotted line.
In the three-epoch example illustrated at the bottom, transition times are put at $t_{1}=7$ and $t_{2}=15$, creating three epochs with substitution processes governed by infinitesimal rate matrices $\mathbf{Q_{1}}$, $\mathbf{Q_{2}}$ and then again $\mathbf{Q_{1}}$, as indicated by the alternating dark and light areas and the dotted lines.
}}
\label{fig:2RateTree}
\end{figure}

In Figure \ref{fig:2RateTree}, we illustrate the tree topology and the transition times defined for both a two-epoch and three-epoch specification.
This tree has 69 tips, is rooted and time-scaled, effectively covering a period of about 18 years. 
For each replicate data set, we simulate $1000$ nucleotide sites under the Hasegawa, Kishino and Yano (HKY) model \citep{hky85}. 
We set the substitution rates and base frequencies to values estimated from the real data set.

In a first scenario, we test whether the epoch model correctly identifies a homogeneous nucleotide substitution process. 
We simulate an alignment evolving under the HKY model with a $\kappa$ parameter (the transition-transversion bias) value of $1.0$ for the whole timespan of the tree. 
In the analyses of replicate data, we specify a boundary time $T_1=7.0$ (7 years before the most recent sampling date) creating $M=3$ ordered boundaries with $S=2$ substitution processes governing character changes between them.
We then run an MCMC chain, starting from a randomly generated tree topology assuming proper log-normal priors on parameters $\kappa_{1}$ and $\kappa_{2}$.
We repeat the simulation and inference process $100$ times and report estimator coverage, bias (mean signed difference) and mean squared error (MSE) in Table \ref{tab:Sim}.
The MSE quantifies the amount by which the estimator differs from the true value of the quantity being estimated.
Estimator coverage reflects the probability that the true value from which the data derive falls within the model estimated nominal 
credible interval and hence predicts the performance of the methods across a wide set of data sets.
While Bayesian credible intervals do not need to yield nominal coverage, we still obtain coverages of 96\% and 98\% for $\kappa_{1}$ and $\kappa_{2}$ respectively.

In a second simulation scenario, we consider a heterogeneous substitution process in which the recent substitution history (more recent than $T_{1}=7.0$) is governed by an HKY model with $\kappa = 1$, and alters to an HKY model with $\kappa = 10.0$ beyond that boundary time. 
By analyzing 100 simulation replicates generated under these settings, we arrive at a coverage of 98\% for $\kappa_{1}$ and 96\% for $\kappa_{2}$.

In a third nucleotide simulation scenario, we consider $S=3$ epochs, where before time $T_{1}=7.0$ substitutions occur under an HKY model with a $\kappa$ value of $1.0$, between $T_{1}$ and $T_{2}=15.0$ under an HKY model with $\kappa=10.0$ and after $T_{2}$ again under an HKY model with $\kappa=1.0$. 
The resulting coverages are 95\%, 95\% and 89\% for $\kappa_{1}$, $\kappa_{2}$ and $\kappa_{3}$ respectively. 
As we introduce more epochs, we observe a concomitant increase in MSE.
This can be expected as  partitioning the time into more intervals will typically leave corresponding epochs less informed as less branch length is located in each epoch. 
For the same value of $\kappa = 1$ in the three-epoch model, the MSE is somewhat higher for the oldest epoch (0.071) compared to the most recent epoch (0.021), which is also in line with more branch length informing the latter (Figure \ref{fig:2RateTree}).

%---SIMULATION RESULTS---%
\begin{sidewaystable}[h]
\medskip
\begin{minipage}{\textwidth} 
\begin{center}
\caption{ \footnotesize {\bf Estimator performance for simulated data sets.} 
The table lists the parameter values used to generate data in first major column and coverage of their estimates, along with measures of variance and bias, in the second major column.
Consecutive rows present the results for the first, second and third nucleotide model simulation for dated-tip samples and the third nucleotide model simulation for contemporaneous sequences (ultrametric tree), followed by the  the results of first, second and third codon model simulation for dated-tip samples and the third codon model simulation for contemporaneous sequences. 
}
\footnotesize{
\begin{tabular}{cccccccccccccc}
\hline 
\multicolumn{2}{c}{\textbf{Simulated}} &  & \multicolumn{11}{c}{\textbf{Estimated}}\tabularnewline
 &  &  & coverage & MSD\footnote{Mean Signed Difference} & MSE\footnote{Mean Squared Error} &  & coverage & MSD & MSE &  & coverage & MSD & MSE\tabularnewline
\hline 
\multicolumn{2}{c}{nucleotides} &  &  & $\kappa_{1}$\footnote{HKY model's transition-transversion bias parameters} &  &  &  & $\kappa_{2}$ &  &  &  & $\kappa_{3}$ & \tabularnewline
\cmidrule(lr){1-2}
\cmidrule(lr){4-6}
\cmidrule(lr){8-10}
\cmidrule(lr){12-14}
 & $\kappa_{1}=1$ &  & 0.96 & 0.068 & 1.005 &  & 0.98 & 0.283 & 1.097 &  & - & - & -\tabularnewline
dated tips & $\kappa_{1}=1$, $\kappa_{2}=10$ &  & 0.98 & 0.007 & 0.008 &  & 0.96 & 0.446 & 1.551 &  & - & - & -\tabularnewline
 & $\kappa_{1}=1$, $\kappa_{2}=10$, $\kappa_{3}=1$ &  & 0.95 & -0.009 & 0.021 &  & 0.95 & 1.166 & 17.614 &  & 0.89 & 0.033 & 0.071\tabularnewline
contemporaneous & $\kappa_{1}=1$, $\kappa_{2}=10$, $\kappa_{3}=1$ &  & 0.96 & -0.002 & 0.002 &  & 0.95 & 0.946 & 3.723 &  & 0.92 & 0.009 & 0.061\tabularnewline
\hline 
\multicolumn{2}{c}{codon} &  &  & $\omega_{1}$\footnote{Yang codon model's non-synonymous to synonymous substitution rate ratio} &  &  &  & $\omega_{2}$ &  &  &  & $\omega_{3}$ & \tabularnewline
\cmidrule(lr){1-2}
\cmidrule(lr){4-6}
\cmidrule(lr){8-10}
\cmidrule(lr){12-14}
 & $\omega_{1}=1$ &  & 0.94 & -0.007 & 0.048 &  & 0.93 & 0.014 & 0.53 &  & - & - & -\tabularnewline
dated tips & $\omega_{1}=0.1$, $\omega_{2}=1$ &  & 0.90 & 0.003 & 0.001 &  & 0.93 & 0.011 & 0.054 &  & - & - & -\tabularnewline
 & $\omega_{1}=0.1$, $\omega_{2}=1$, $\omega_{3}=0.1$ &  & 0.92 & 0.002 & 0.001 &  & 0.89 & 0.096 & 0.274 &  & 0.96 & 0.10 & 0.02\tabularnewline
contemporaneous & $\omega_{1}=0.1$, $\omega_{2}=1$, $\omega_{3}=0.1$ &  & 0.93 & -0.002 & 0.001 &  & 0.96 & 0.067 & 0.051 &  & 0.95 & 0.003 & 0.002\tabularnewline
\end{tabular}
} % END: footnotesize
\label{tab:Sim}
\end{center}
\end{minipage}
\end{sidewaystable}
      
Epoch models are not restricted to nucleotide models; they can also relax time-homogeneity in full codon substitution models, such as the Goldman-Yang (GY94) codon model \citep{Goldman1994}.
We here examine the performance of such codon models in an epoch setting. 
As before, we first test a homogeneous substitution scenario and check whether the model is able to recover homogeneous values for the $\omega$ parameters across epochs. 
To this end, we simulate $500$ nucleotide triplets under the GY94 codon model with an $\omega$ parameter value of $1.0$. 
Performing 100 simulation replicates yields a coverage of 94\% for $\omega_{1}$ and 93\% for $\omega_{2}$. 
To asses the coverage in a heterogeneous codon substitution scenario, we set the true values to $\omega_{1}=0.1$ and $\omega_{2}=1.0$, with a transition time $T_{1}=7.0$ between the epochs, which results in a coverage of 90\% and 93\% for $\omega_{1}$ and $\omega_{2}$ respectively.

Analogous to the nucleotide simulations, we also asses the epoch model performance when the data are simulated over three heterogeneous epochs, with sequences evolving under the GY94 codon model with $\omega_{1}=0.1$ before $T_{1}=7.0$, then with $\omega_{2}=1.0$ and after time $T_{2}=15.0$ with $\omega_{3}=0.1$. 
We obtain a coverage of 92\%, 89\% and 96\% for $\omega_{1}$, $\omega_{2}$ and $\omega_{3}$ respectively.
Also in this case the MSE is higher for the oldest epoch compared to the most recent epoch.

For both nucleotide and codon models we also explore how well epoch parameters can be recovered from contemporaneous sequence data, 
% so 
without sequences sampled throughout the past epochs.
To this end we set all sampling dates to time $t = 0$, effectively transforming the tree topology to be ultrametric (all tips at equal distance from the root, Supplementary Figure 1).
We list the results for these simulations under the rows labelled as `contemporaneous' in Table \ref{tab:Sim}. 
The resulting coverages for contemporaneously sampled sequences are 96\%, 95\% and 92\% for $\kappa_{1}$, $\kappa_{2}$ and $\kappa_{3}$ respectively and 93\%, 96\% and 95\% for $\omega_{1}$, $\omega_{2}$ and $\omega_{3}$  respectively.
We note that the MSE is generally lower for estimates produced for the contemporaneous data because the ultrametric transformation implies that more branches inform the epochs (Supplementary Figure 1).

%%%%%%%%%%%%%%%%%%%%%%%%
%---SHANKARAPPA DATA---%
%%%%%%%%%%%%%%%%%%%%%%%%
\subsection{Within-host HIV selection dynamics}

We re-analyze within-host HIV-1 sequence data from eight patients extensively sampled throughout infection starting close to the time of seroconversion \citep{Shankarappa99}.
These patients have previously been classified as moderate or slow progressors based on progression time, or the time it takes for CD4+ T cell counts to drop below 200 cells/$\mu$l \citep{williamson03}.
The data consist of \textit{env} C2V5 sequences collected over a 6 to 13.7 year period with an average of 12 time points per patient (see supplementary material). 
The original investigation of HIV-1 diversity and divergence over time in these patients reveals a consistent pattern of divergence stabilization at late-stage infection \citep{Shankarappa99}.
This has led to two different hypotheses that may explain these patterns.
The immune relaxation hypothesis posits that the damaged immune system during the symptomatic stage leads to reduced selection pressure on the virus, which relaxes the need for fixing immune escape mutations in the viral population.
The cellular exhaustion hypothesis, on the other hand, states that the decreased target cell availability  in late-stage infection provides less opportunity for viral replication. While the former only impacts non-synonymous changes, the latter is expected to reduce both synonymous and non-synonymous rates of substitutions.

To distinguish between these hypotheses, we ask whether $\omega$
decreases at late-stage infection, as defined by the progression time for each patient. 
This rate ratio is an explicit parameter of the GY94 codon substitution model \citep{Goldman1994}, which we can extend with an epoch specification.  
For each patient, we compare a standard homogeneous model to a two-epoch specification with a separate GY94 model before and after boundary time $T_{1}$ set to progression time for that patient.
We exclude patient 11 from the original study because no sequence data are available after progression time for this patient \citep{Shankarappa99}.
The two-epoch discretization allows estimating a separate $\omega$ parameter for the two infection stages in each patient, with $\omega_{2}$ denoting the $d_N/d_S$ ratio before progression and $\omega_{1}$ denoting the same parameter after progression.

%---SHANKARAPPA ANALYSIS RESULTS HPD---%
\begin{figure}[H]
\begin{center}
\definecolor{shadecolor}{rgb}{0.969, 0.969, 0.969}\color{fgcolor}\includegraphics[scale=0.75]{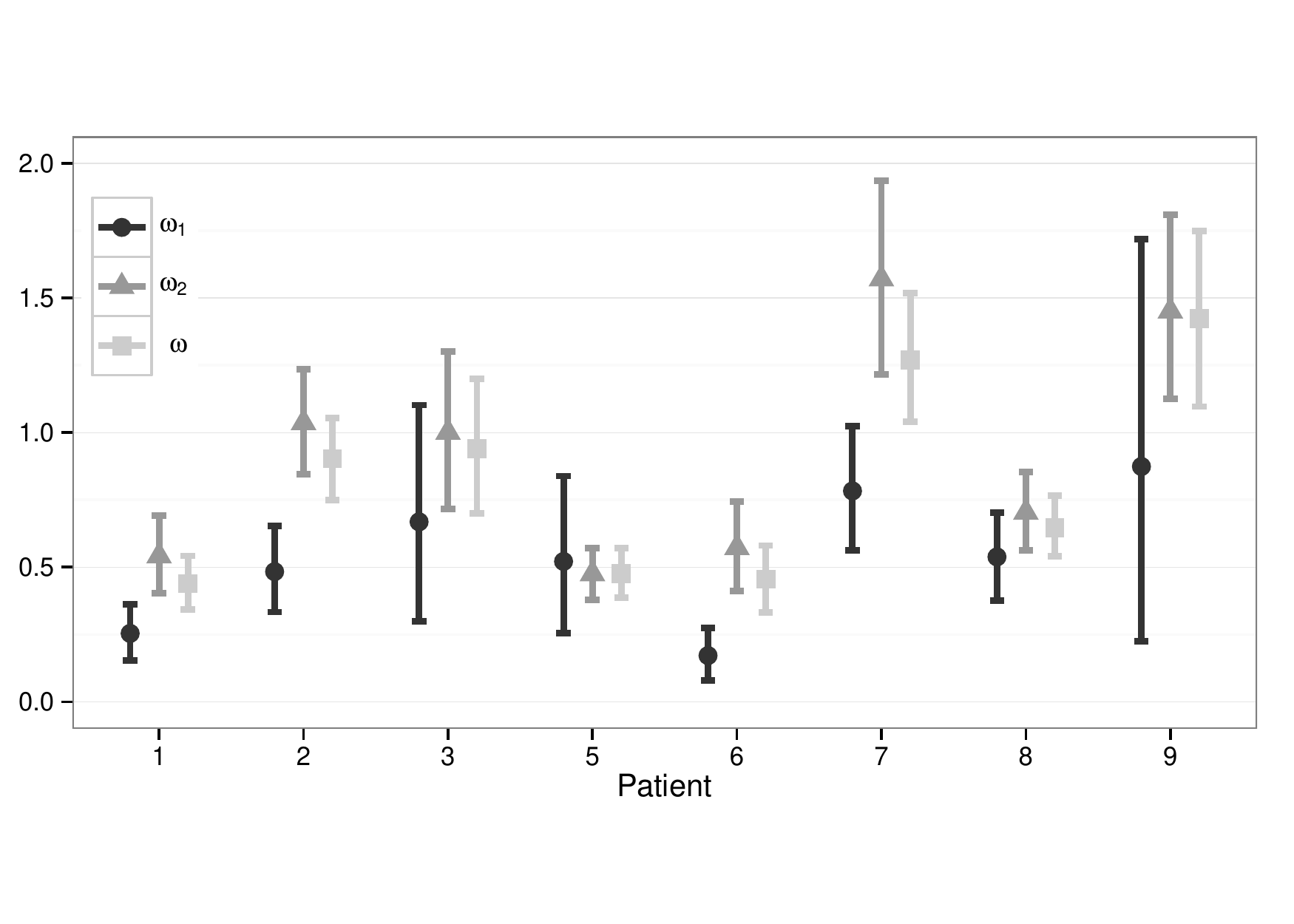} 
\end{center}
\caption{{ \footnotesize {\bf Estimates of $d_N/d_S$ ratio for within-host HIV analyses.} Vertical lines represent 95\% highest posterior density intervals for the $d_N/d_S$ ratio estimates. 
Parameter $\omega$ is estimated under the homogeneous model, while $\omega_{1}$ and $\omega_{2}$ are obtained using the epoch model.
}}
\label{fig:shank_hpd}
\end{figure}

Figure \ref{fig:shank_hpd} presents the results for the $\omega$ parameter estimates. 
The $\omega$ estimates indicate a general decrease in $d_N/d_S$ after progression time ($\omega_{1} < \omega_{2}$, Figure \ref{fig:shank_hpd}).
The most pronounced differences in $\omega$ before and after progression time can be observed for patients 1, 2, 6 and 7.
For patients 2, 3, 7 and 9, the drop in mean $\omega$ estimates suggests a shift in neutral or even positive selection ($\omega_{2} \geq 1$) to negative selection ($\omega_{1} < 1 $).
The homogeneous $\omega$ estimate is generally closer to $\omega_{2}$, which can be expected because most evolutionary history takes place prior to progression time. 

%---SHANKARAPPA ANALYSIS RESULTS BF---%
\begin{table}[H]
\medskip
\begin{minipage}{\textwidth} 
\begin{center}
\footnotesize{
\begin{tabular}%{ccc}
{@{}l*{2}{D{.}{.}{7}}@{}}
\hline 
\textbf{Patient} & \textbf{Posterior probability} & \textbf{log Bayes factor}\tabularnewline
\hline 
patient 1 & $0.999$ & $7.418$\tabularnewline
patient 2 & >$0.999$ & $9.602$\tabularnewline
patient 3 & $0.898$ & $2.174$\tabularnewline
patient 5 & $0.430$ & $-0.282$\tabularnewline
patient 6 & >$0.999$ & $9.210$\tabularnewline
patient 7 & >$0.999$ & $8.112$\tabularnewline
patient 8 & $0.933$ & $2.627$\tabularnewline
patient 9 & $0.895$ & $2.142$\tabularnewline
\hline 
\textbf{Joint evidence:} & $0.894$ & $2.14$ \tabularnewline
\end{tabular}
} % END: footnotesize
\caption{{ \footnotesize {\bf Bayes factor test for decreased selection after progression.} We report the posterior probability that $\omega_{1} < \omega_{2}$ and the corresponding Bayes factor against the alternative that $\omega_{1} \ge \omega_{2}$.
} }
\label{tab:shank_bf}
\end{center}
\end{minipage}
\end{table}

Despite the observation that the Bayesian credible intervals for patient 1, 2, 6 and 7 estimates do not overlap, this does not provide a formal test to evaluate their differences.
Therefore, we conduct a Bayes factor (BF) test \citep{suchard2005models} that expresses the posterior odds over the prior odds that $\omega_{1} < \omega_{2}$ for the individual analyses of each patient.
To determine the posterior odds, we note that the MCMC sample average of an indicator function that the parameter values fall within one competing model space converges to the posterior probability of that model.
The prior odds in our case is simply 1.
The log Bayes factors listed in Table \ref{tab:shank_bf} suggest generally strong evidence for a declining selective pressure after progression, with one notable exception for patient 5.
We also provide a Bayes factor that summarizes the joint evidence for $\omega_{1} < \omega_{2}$, which suggest an overall support in favor of the immune relaxation hypothesis (log BF = 2.14), in accordance with previous findings suggesting a general decrease in non-synonymous divergence at late-stage infection \citep{williamson2005, Lemey07}.

%%%%%%%%%%%%%%%%%%%%%%
%---PHYLOGEOGRAPHY---%
%%%%%%%%%%%%%%%%%%%%%%
\subsection{Seasonal circulation dynamics of human influenza A}

In a second application of the epoch model, we focus on discrete diffusion processes to infer spatio-temporal history from viral gene sequences.
This type of phylogeographic inference, where the sampling locations are considered as discrete geographic traits, has gained popularity in recent years, at least partly because of a flexible and efficient Bayesian implementation that connects dispersal dynamics to sequence evolution in time-measured phylogenies \citep{Lemey2009}.
Recently \citet{Bahl14112011} have applied this Bayesian inference framework to investigate the circulation dynamics of global influenza A H3N2 through time.
Since the authors were interested in capturing the heterogeneity in these dynamics over successive seasonal epidemics between 2003 and 2006, they consider discrete traits that are the product of sampling location and sampling time (epidemic season).
Not only does this discretization by sampling time seem counterintuitive for a model that emits discrete outcomes as a continuous function of time, it also considerably increases the dimensionality of the CTMC rate matrix and thus the number of parameters to inform by the sparse spatial data.

\begin{figure}[H]
\begin{center}
\includegraphics[scale=0.5]{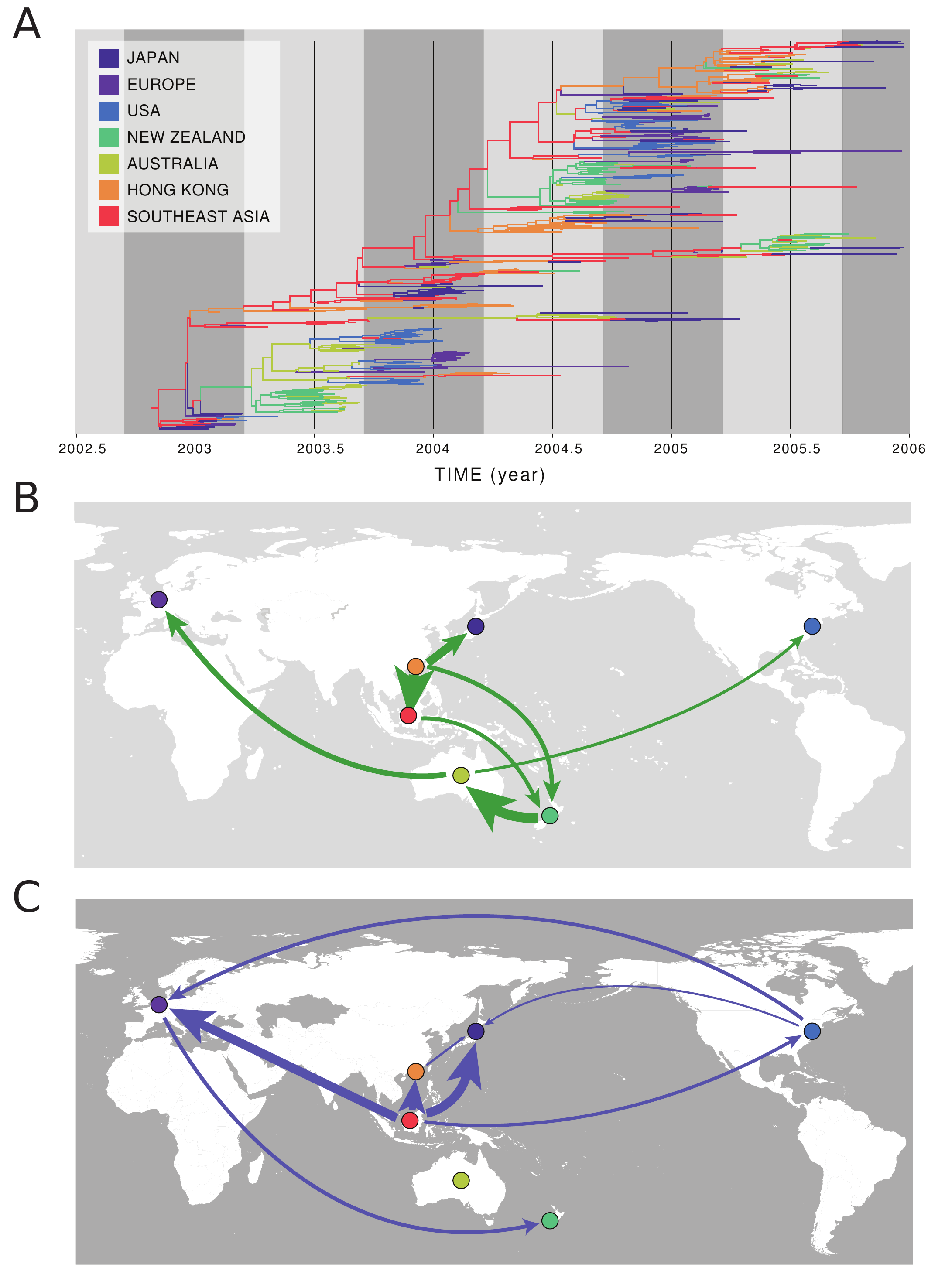} 
\end{center}
\caption{{ \footnotesize {\bf A two-epoch phylogeographic model applied to seasonal influenza H3N2.} 
A. Maximum clade credibility (MCC) tree with branches colored according to modal discrete location states at each node. The grey time intervals represent the epoch model with a single discrete rate matrix shared across northern hemisphere spring and summer (light grey) time intervals and another rate matrix shared across the northern hemisphere autumn and winter (dark grey) time intervals. B. Diffusion rates supported by a Bayes factor $>20$ for spring and summer epoch intervals. The width of the arrows reflects the magnitude of the Bayes factor support. C. Diffusion rates supported by a Bayes factor $>20$ for autumn and winter epoch intervals.
} }
\label{fig:2EpochFlu}
\end{figure}

Here, we explore epoch time-discretization as a more appropriate alternative to detect temporal heterogeneity in influenza dispersal.
We revisit the \citet{Bahl14112011} data set that consists of 525 influenza A H3N2 hemagglutinin sequences sampled from  Australia, Europe, Japan, New York, New Zealand, Southeast Asia and Hong Kong ($n=75$ each) from 2003 to 2006.
In a first epoch model extension of the discrete phylogeographic approach, we specify alternating epochs for the time intervals encompassing northern hemisphere spring and summer and the time intervals encompassing northern hemisphere autumn and winter.
The discrete diffusion parameters are shared across rate matrices for the spring and summer epochs as well as for the autumn and winter epochs, effectively producing two rate matrices compared to a single matrix for the homogeneous model. Figure \ref{fig:2EpochFlu} schematically represents this $S=7$ epoch parametrization. 
Following \cite{KassRaftery95} we report rates that yield a Bayes factor support interpreted as `strong evidence'.

%---FLU DATA RESULTS MAP---%

We apply a Bayesian stochastic search variable selection (BSSVS) procedure to identify the best supported diffusion rates within each epoch using a Bayes factor test, as available in the SPREAD software \citep{Bielejec11092011}. Rates yielding a Bayes factor over 20 are represented in Figure \ref{fig:2EpochFlu}A\&B for the spring and summer epoch and autumn and winter epoch respectively.
This suggests seasonal dynamics with spring and summer circulation to a large extent mirroring autumn and winter circulation.
The spring and summer epoch appears to be dominated by circulation from Southeast Asia and Hong Kong to the Southern hemisphere (New Zealand), circulation within the Southern hemisphere and also circulation from the Southern to the Northern hemisphere.
During the autumn and winter epoch on the other hand, we infer mostly circulation from Southeast Asia to the Northern hemisphere, circulation within the Northern hemisphere and occasional circulation from the Northern to the Southern hemisphere,

%---FLU DATA RESULTS PS SS---%
\begin{table}[H]
\begin{minipage}{\textwidth} 
\begin{center}
\footnotesize{
\begin{tabular}{ccc}
\hline 
\multirow{2}{*}{\textbf{Model}} & \multicolumn{2}{c}{\textbf{Marginal likelihood}}\tabularnewline
 & PS\footnote{Path Sampling} & SS\footnote{Stepping Stone Sampling}\tabularnewline
\hline 
homogeneous & -827.29 & -825.07\tabularnewline
7-epoch & -806.36  & -803.40 \tabularnewline
14-epoch & -798.77  & -795.63 \tabularnewline
\end{tabular}
} % END: footnotesize
\caption{{ \footnotesize {\bf Marginal likelihood estimates.} Comparison in terms of model fit between a homogeneous model, an epoch model with time discretized into $S=7$ epochs alternating between 2 different rate matrices and an epoch model with time discretized into $S=14$ epochs, alternating between 4 separate rate matrices.
} }
\label{tab:flu_ps}
\end{center}
\end{minipage}
\end{table}

To evaluate the improvement of explicitly modeling these largely opposing dynamics, we compared model fit with a homogeneous model using path sampling and stepping-stone sampling, two reliable estimators of marginal likelihood \citep{Baele07032012}. 
Proper priors were used for all parameters during the various analyses, as well as the model selection, since such priors have been shown to be essential when performing marginal likelihood estimation \citep{Baele22102012}.
The results of the model comparison are listed in Table~\ref{tab:flu_ps} and provide evidence for the two-epoch model outperforming the homogeneous model.
When we further extend our phylogeographic epoch time-discretization to four epochs, modeling separate dynamics for each individual season, we observe additional improvements in terms of marginal likelihoods but with diminishing returns with respect to the two-epoch vs. homogeneous comparison (see Supplementary Material for a visual summary of well supported circulation rates in each season).

%%%%%%%%%%%%%%%%%%%%%%%%%
%---RUN TIME ANALYSIS---%
%%%%%%%%%%%%%%%%%%%%%%%%%
\subsection{Run time analysis}

To investigate the speed increase offered by a parallel execution of different tasks, we measure the speed-up relative to serial execution of both update and convolution operations for the analyses of two empirical data sets.
We perform our analyses on a standard desktop PC equipped with Intel 2$^{\text{nd}}$ Gen Core i7 2 i7-2600K / 3.9 GHz CPU and 16 GB of DDR3 SDRAM along with a NVIDIA GeForce GTX 590 card, carrying two GPU devices with a total of 1024 CUDA cores, running at 1215 MHz and 1536MB of memory per GPU. 

%---STACK SIZE TIMING---%
\begin{figure}[H]
\begin{center}
\includegraphics[width=5in]{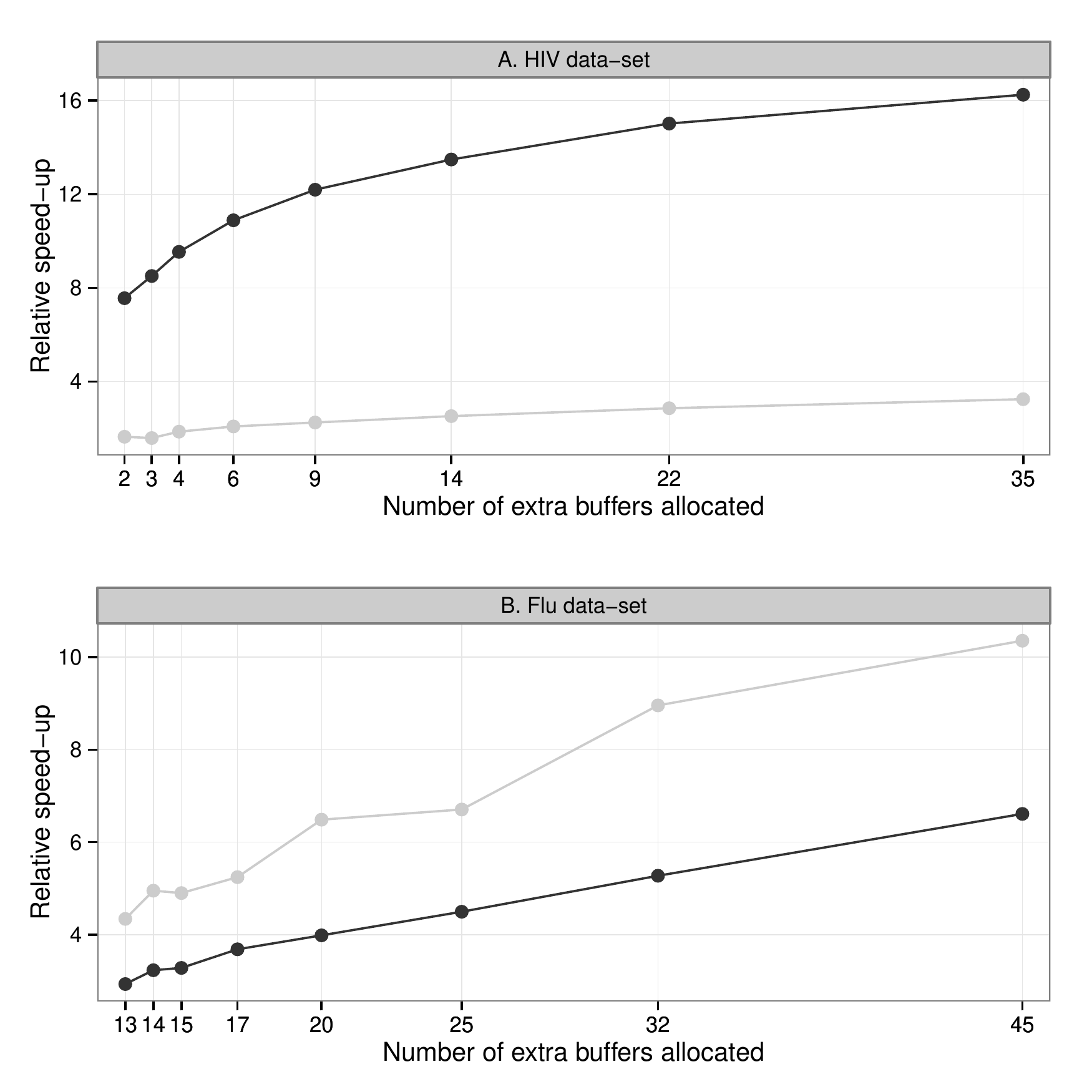} 
\end{center}
\caption{{ \footnotesize {\bf Relative speed-ups as a function the number of extra buffers allocated for two real data examples.}
Speed-ups are reported relative to the serial execution, using system time spent sampling 1,000,000 states from the posterior distribution using an MCMC sampler. 
Black dots denote update transition matrices operations, grey dots denote convolve operations.
A. Speed-ups for a flu data set (522 taxa, discrete geographic diffusion process between $K=7$ states, time discretized into $S=13$ epochs, see {\it{Seasonal circulation dynamics of human influenza A}}). 
B. Speed-ups for a patient 3 from HIV data set (109 taxa, 516 sites, state-space size of $K=61$ codons, $S=2$ epochs, see {\it{Within-host HIV selection dynamics}}). 
} }
\label{fig:StackSizeTiming}
\end{figure}

Figure \ref{fig:StackSizeTiming} presents the GPU speed-ups induced by different choices of the number of extra buffers, which in turn determine the queue size for update and convolve operations. 
The leftmost points correspond to the minimum number of extra buffers we can allocate; speed-ups are reported relative to the serial execution of the convolve and update steps on a CPU device. We perform our analyses with double floating point precision on both the GPU and CPU.
We note that the speed-ups between data sets depend not only on the number of epoch transition times, but also importantly on the actual number of convolution operations performed during the tree topology traversal and this is tree-dependent. 
The convolution step for the influenza example emerges as the most time consuming part of the routine (Figure \ref{fig:StackSizeTiming} AB), as many branches intersect multiple epochs, breeding multiple calls to the \emph{ConvolveTransitionMatrices} BEAGLE kernel.  
For the codon-based HIV-1 example (Figure \ref{fig:StackSizeTiming} A) a large state space alongside with a modest number of branches intersecting epoch transition boundaries provide the main reason why we witness a significant amount of time being spent computing finite-time transition probabilities. 
As for all the HIV-1 data sets we also compare sequential and parallel absolute timings for an `epochized' codon model, measured by the time it takes MCMC chain to sample 1 million posterior samples (see the Supplementary Information).

\section{Discussion}

\citet{Goode2008} demonstrate that a change in evolutionary pattern affecting all individuals of a population can be modeled by specifying different substitution models across different time intervals rather than lineage-specific substitution models.
Here, we extend this approach and further demonstrate how epoch modeling can uncover temporal heterogeneity in discrete character evolution in phylogenetic histories.
We are mainly interested in heterogeneity resulting from variation in the relative intensities of substitutions across time and not heterogeneity induced by non-stationarity.
We embed the epoch model in a Bayesian phylogenetic framework that focuses entirely on time-measured trees and integrates over all plausible evolutionary histories for the observed sequence data.
Our simulations show that the model is able to recover different scenarios of heterogeneity under different substitution models, but epoch parameters can also reflect an underlying process that is in fact homogeneous, thus avoiding false positives (see Table \ref{tab:Sim}).
Following \citet{Goode2008}, we primarily focus on time-stamped sequence data from rapidly evolving pathogens for which the ecological and evolutionary dynamics occur on the same time scale and potentially interact.
However, our simulation study demonstrates that the model is at least equally capable to capture sequence substitution heterogeneity through time from contemporaneous data, and the amount of evolutionary history (branch length) in an epoch determines how well epoch parameters can be recovered.

We apply our model both in the context of sequence evolution and spatial dispersal dynamics.
For the former, we focus on within-host HIV-1 evolution and explicitly test different hypotheses that explain the stabilization in sequence divergence in late-stage infection \citep{Shankarappa99}.
This phenomenon has been attributed to weakened selection pressure (immune relaxation) or to a decrease in average viral replication rate (cellular exhaustion) \citep{williamson2005}.
Various studies have attempted to distinguish between both scenarios by contrasting the accumulation of non-synonymous and synonymous  substitutions using different methodologies \citep{williamson2005,Lemey07,Lee2008}.
While two studies provide strong support for the immune relaxation hypothesis using different methodologies \citep{williamson2005,Lemey07}, \citet{Lee2008} suggest that both synonymous and non-synonymous evolutionary rates decline as disease progresses.
Here, we explicitly model a change in the $d_N/d_S$ ratio in codon substitution models while integrating over the underlying within-host HIV-1 phylogeny, and formally evaluate the support for a decrease in $d_N/d_S$ using Bayes factors. 
This demonstrates strong overall support in favour of the immune relaxation hypothesis.

Our second application exemplifies  the use of epoch time-discretization in phylogeographic inferences that consider sampling locations as discrete traits.
In particular, we demonstrate how epoch modeling can capture seasonality in human influenza A H3N2 circulation dynamics, without the need to complicate location traits with sampling time.
Based on a data set previously analyzed by \citet{Bahl14112011}, we infer different epidemiological connections between the northern hemisphere spring-summer epoch and the autumn-winter epoch.
In both cases Southeast Asia (and Hong Kong) appear to play a central role in seeding the seasonal epidemics in the different hemispheres (Figure \ref{fig:2EpochFlu}). 
However, we remain cautious in interpreting the support for diffusion rates in the context of source-sink dynamics because strong evidence for such a rate being non-zero does not necessarily imply that the diffusion rate itself is high \citep{Faria13}.
For example, the connection we identify between Europe and New Zealand during the northern hemisphere autumn and winter epidemic may represent a few introductions into New Zealand without extensive onwards transmission.
So, it remains difficult to assess which rates govern potential source-sink dynamics.

The specification of two alternating epochs yields a better model fit than a homogeneous model while providing a more parsimonious parametrization compared to the use of discrete traits that are based on both sampling location and epidemic season as in \citet{Bahl14112011}.
Model fit differences are more readily detected in this application because an additional epoch adds an entirely new set of parameters.
Incorporating more epochs further increased model fit (Table \ref{tab:flu_ps}), albeit with diminishing returns, but it is clear that there are limits to the flexibility that can be incorporated in phylogeographic reconstructions, which represent inherently data-sparse inferences.
In this respect, it is interesting to note that approaches are available to share information across epochs while still allowing for the detection of differences among them \citep{suchard2003hierarchical}.
This can be achieved by specifying hierarchical priors, both on standard rate parameters \citep{Edo-Matas2011} as well as, more recently, on the rate indicators in a BSSVS procedure \citep{cybis2013graph}.
Assuming that phylogeography represents the major application for the epoch model, further research is needed to explore these approaches as well as other sparse parameterizations of discrete dispersal processes.

The two data sets we examine represent examples with clear \textit{prior} hypotheses that correspond to fixed transition times.
It may also be of interest to apply the model when the transition times are unknown. 
Although it would be straightforward to estimate the time of a fixed number of epoch transition in our MCMC framework, estimating the number of epochs may be more challenging.
However, our previous experience with change-point processes in phylogenetics \citep{suchard2003inferring} suggests that it should be possible to introduce prior distributions over these quantities and jointly infer them when uncertainty remains in their specification.

By adding an additional layer of complexity to our evolutionary models and inference framework, we further increase the computational demands in a field that is already computationally intensive. 
Our Bayesian approach integrates over all possible evolutionary scenarios, which is challenging for a large number of sequences even when specifying the simplest of evolutionary models.
To mitigate the additional burden imposed by our epoch time-discretization and the operations involved, we have implemented our model in the high-performance BEAGLE library allowing us to perform the calculations on GPU architectures.
Although this has proven extremely useful, in particular for large state-space models such as codon substitution models \citep{Suchard2009}, and is also demonstrated by comparing serial and parallel run time estimates for the Shankarappa data sets, more research is needed to further stretch the limits of practical computational restrictions.
 
In summary, our work has extended the phylodynamic framework with a model that is capable of quantifying and testing temporal heterogeneity in discrete state transition processes, which is proving useful to detect changing selective dynamics in rapidly evolving viral populations as well as fluctuations in historical circulation dynamics.

\section{Software availability}

BEAST (Bayesian Evolutionary Analysis by Sampling Trees) source code is freely available at \url{http://code.google.com/p/beast-mcmc/} under the terms of GNU LGPL license. Compiled, ready-to-use binaries targeting major platforms can be obtained from \url{http://beast.bio.ed.ac.uk}. 
The BEAGLE (Broad-platform Evolutionary Analysis General Likelihood Evaluator) library is free, open-source software licensed under the GNU LGPL. Both the source code and binary installers are available from \url{www.code.google.com/p/beagle-lib/}. 
SPREAD (Spatial Phylogenetic Reconstruction of Evolutionary Dynamics) is licensed under the GNU Lesser GPL, and its source code is freely available at \url{https://github.com/phylogeography/SPREAD}. Compiled, runnable packages are hosted at \url{http://www.phylogeography.org/SPREAD.html}. 

\section{Funding}

The research leading to these results has received funding from the European Research Council under the European Community's Seventh Framework Programme (FP7/2007-2013) under Grant Agreement no. 278433-PREDEMICS and ERC Grant agreement no. 260864,
the Wellcome Trust (grant no. 092807),
the National Science Foundation (DMS 1264153)
and the National Institutes of Health (R01 HG006139). The National Evolutionary Synthesis Center (NESCent) catalyzed this collaboration through a working group (NSF EF-0423641).

\section{Acknowledgments}

We wish to thank Justin Bahl for providing the influenza data set and Nuno Faria, Nidia Trovao and Bram Vrancken for their critical insight and testing of the software. 
 
% \bibliography{epochrefs}

%%%%%%%%%%%%%%%%%%%%%%%%%%%%%%%%%%%
%---BEAST/BEAGLE IMPLEMENTATION---%
%%%%%%%%%%%%%%%%%%%%%%%%%%%%%%%%%%%
\subsection*{Appendix A: Fine-grain parallel implementation}

In this Appendix, we describe the fine-grain parallel implementation of the epoch model in the BEAGLE library \citep{ayres2012beagle}.
Pursuing the single instruction, multiple data (SIMD) programming paradigm, GPUs use a thread management system 
that executes many, possibly millions of lightweight threads at a relatively modest speed, rather than a limited number of threads very rapidly on multiple-core CPUs.
Concurrent threads are grouped into blocks and blocks are organized as a two-dimensional grid, with their sizes specified upon invocation of the function executed on the device, referred to as a kernel. 
On the current NVIDIA GPU architecture, a grid can consist of thousands of thread blocks, each containing hundreds of threads, thus resulting in a fine-grain data parallelism.
Threads on the grid share a common memory address space referred to as the global memory, while threads belonging to the same block share a very fast user-managed cache, known as shared memory.  
Global memory access is the slowest, but it also comprises the largest memory region, measured in GByte orders of magnitude on current architectures, whereas shared memory size is much more limited and measured in KBytes.
The BEAGLE API allows
allocating memory for the subsequent tree-pruning operations, as well as specifying key features of the character data and substitution model.  
BEAGLE efficiently performs numerical calculation of transition probability matrices from eigen decompositions of the rate matrices $\mathbf{Q}_{s}$ for arbitrary rate scalar $r$ and edge length $t$.

This approach is highly efficient as the repeated evaluations of $\exp\left(rt\mathbf{Q}_s\right)$ for different $r$ and $t$ reuse the same decomposition of the rate matrix.
Calculating the diagonal matrix of the resulting decomposition of $\exp\left(rt\mathbf{Q}_s\right)$ only involves scaling and exponentiating the eigenvalues of $\mathbf{Q}_s$, effectively reducing the time-consuming calculations to dense matrix-vector-matrix multiplication.
BEAGLE achieves both 
fine-grain parallelism here via massive SIMD reduction of the matrix-vector-matrix multiplication and
coarse-grain parallelism by computing transition matrices across branches in parallel.  

\newcommand{\blockSize}{\ensuremath{B}}

Discretizing the time component introduces additional layers of complexity to the likelihood computations. 
Multiple rate matrices determine the substitution process along branches that traverse different epochs, requiring piecewise matrix convolution operations to arrive at the common process operating on such branches.  
We pursue fine-grain parallelism for dense matrix-matrix multiplication by breaking up the problem into smaller sub-problems that fit into the limited amount of shared memory. 
This involves multiplication within a block matrix in which, following the notation from Figure \ref{fig:collapsing}, each one of the $K \times K$ matrices $\mathbf{P}_{1}(t_u, T_1, r_v)$, $\mathbf{P}_{2}(T_1, t_v, r_v)$ and the product matrix $\mathbf{P}(t_u, t_v, r_v)$ are partitioned into $\blockSize \times \blockSize$ sub-matrices. 
We choose the sub-matrix dimensions such that $\blockSize$ is a multiple of $K$ when possible to avoid thread-divergence. 
Every entry in $\mathbf{P}(t_u, t_v, r_v)$ is therefore computed as
\begin{equation}
\left\{ \mathbf{P}(t_{u},t_{v},r_{v})\right\} _{ij}=\underset{0\leq k\leq
K / \blockSize
}
{\sum}\left\{ \mathbf{P}_{1}(t_{u},T_1,r_{v})\right\} _{ik}\times\left\{ \mathbf{P}_{2}(T_1,t_{v},r_{v})\right\} _{kj} .
\end{equation}
This strategy executes all sub-problems in parallel%
, with each thread computing a dot product of row-vector  
$\left\{ \mathbf{P}_{1}(t_{u},T_1,r_{v})\right\} _{i\cdot}$  
and column vector 
$\left\{ \mathbf{P}_{2}(T_1,t_{v},r_{v})\right\} _{\cdot j}$ 
stored and reused across threads in shared memory to produce thread-specific element 
$\left\{ \mathbf{P}(t_{u},t_{v},r_{v})\right\} _{ij}$ 
that transfers to global memory only once. 

\end{document}